\documentclass[11pt]{article}
\usepackage{graphicx}
\usepackage{cite}
\usepackage{amssymb}
\usepackage{epstopdf}
\usepackage{hep}
\usepackage{feynmp}
\usepackage{multicol}
\usepackage{simplewick}
\usepackage{appendix}
\usepackage{subfigure}
\usepackage{color}
\usepackage{ulem}
\allowdisplaybreaks

\makeatletter
\def\slash#1{{\mathpalette\c@ncel{#1}}} 
\makeatother  

\newcommand{\sh}{\hat{s}}
\def\th{\hat{t}}
\newcommand{\uh}{\hat{u}}

\newcommand{\eps}{\epsilon}

\newcommand{\pup}{p^{\uparrow}}

\newcommand{\Sp}{S_{\perp}}
\newcommand{\Gt}{\widetilde{G}}

\newcommand{\sigmah}{\hat{\sigma}}

\newcommand{\St}{\tilde{S}}

\newcommand{\vkT}{\vec{k}_T}

\newcommand{\nn}{\nonumber}

\newcommand{\bfc}{\begin{figure}\begin{center}}
\newcommand{\efc}{\end{center}\end{figure}}

\newcommand{\fig}[2]{\scalebox{#1}{\includegraphics{#2}}}

\newcommand{\Slash}[1]{\ooalign{\hfil/\hfil\crcr$#1$}}

\newcommand{\Tr}[1]{{\rm Tr}\left[ #1 \right]}
\newcommand{\vb}{\biggl|}
\newcommand{\cl}{\biggr|_{\rm c.l.}}

\newcommand{\psib}{\bar{\psi}}
\newcommand{\del}[2]{\frac{\partial #1}{\partial #2}}

\def\pslash{\rlap/{\mkern-1mu p}}
\def\Sslash{\slash{\mkern-1mu S}}
\def\kslash{\slash{\mkern-1mu k}}
\newcommand\beq{\begin{eqnarray}}
\newcommand\eeq{\end{eqnarray}}
\def\Fbar{\bar{F}}

\begin{document}

\title{{\bf Transverse single-spin asymmetries in {\boldmath $p^\uparrow p \to \gamma X$} from quark-gluon-quark correlations in the proton}}

\author{K.~Kanazawa$^{1}$, Y.~Koike$^{2}$, A.~Metz$^{1}$, and D.~Pitonyak$^{3}$
 \\[0.3cm]
{\normalsize\it $^1$Department of Physics, Barton Hall,
  Temple University, Philadelphia, PA 19122, USA} \\[0.15cm]
{\normalsize\it $^2$Department of Physics, Niigata University,
Ikarashi, Niigata 950-2181, Japan} \\[0.15cm]
{\normalsize\it $^3$RIKEN BNL Research Center,
                 Brookhaven National Laboratory,
                 Upton, New York 11973, USA} 
}

\date{\today}
\maketitle

\begin{abstract}
\noindent
We analyze the transverse single-spin asymmetry in direct photon production from proton-proton collisions, denoted $A_N^\gamma$, within collinear twist-3 factorization.  We provide a calculation of the contribution due to quark-gluon-quark correlations in the unpolarized proton as well as summarize previous studies on those effects in the polarized proton.  Both soft-gluon poles and soft-fermion poles are considered.  From this complete result we then estimate $A_N^\gamma$, including error bands due to uncertainties in the non-perturbative inputs, at kinematics relevant for planned measurements of this observable at the Relativistic Heavy Ion Collider.  We find $A_N^\gamma$ can allow for a ``clean'' extraction of the Qiu-Sterman function, which could lead to a definitive solution to the so-called ``sign mismatch'' crisis.  Since we use the Sivers function extracted from semi-inclusive deep-inelastic scattering to develop our input for the Qiu-Sterman function, this reaction can also make a statement about the process dependence of the Sivers function. 
\end{abstract}

%
%
\section{Introduction} \label{s:intro}

The transverse single-spin asymmetry (TSSA) $A_N$ has been studied since the mid-1970s.  Large effects first measured in polarized lambda production at FermiLab \cite{Bunce:1976yb} proved difficult to describe in perturbative QCD~\cite{Kane:1978nd}.  In the 1980s it was shown that quark-gluon-quark correlations in the nucleon could lead to substantial TSSAs~\cite{Efremov:1981sh}.  In the 1990s this formalism, known as collinear twist-3 factorization, was worked out for proton-proton collisions in more detail, first for direct photon production \cite{Qiu:1991pp,Qiu:1991wg} and then for pion production \cite{Qiu:1998ia}.  Several other analyses furthered the development of this framework --- see, e.g., \cite{Eguchi:2006qz,Eguchi:2006mc,Kouvaris:2006zy,Koike:2006qv,Koike:2009ge,Kanazawa:2010au,Kanazawa:2013uia,Metz:2012ct}.  In addition, this theoretical work has been complemented by many experimental measurements of $A_N$ at proton-(anti)proton accelerators over the last two decades \cite{Adams:1991rw,Krueger:1998hz,Adams:2003fx,Adler:2005in,Lee:2007zzh,:2008mi,Adamczyk:2012xd,Bland:2013pkt,Adare:2013ekj}.  Most of this experimental data has come in the form of light-hadron asymmetries $A_N^h$, e.g., $h=\pi$, $K$, $\eta$, with the exception of the jet asymmetry $A_N^{jet}$ recently measured by the A$_N$DY Collaboration~\cite{Bland:2013pkt}.  So far, no measurement of the direct photon asymmetry $A_N^\gamma$ has been performed, although there are plans to carry out such experiments at the Relativistic Heavy Ion Collider (RHIC) by both the PHENIX Collaboration \cite{PHENIX:BeamUse} and the STAR Collaboration \cite{STAR:BeamUse}.

At this stage we feel it is important to understand the analytical structure of the single-spin dependent cross section for these aforementioned collinear twist-3 observables.  For a general process $A^\uparrow + B \rightarrow C + X$, this cross section can be written as the sum of three terms:
\begin{align} 
d\sigma(\vec{S}_{\perp}) &= \,H\otimes f_{a/A(3)}\otimes f_{b/B(2)}\otimes D_{C/c(2)} \nonumber \\*
&+ \,H'\otimes f_{a/A(2)}\otimes f_{b/B(3)}\otimes D_{C/c(2)} \nonumber \\
&+ \,H''\otimes f_{a/A(2)}\otimes f_{b/B(2)}\otimes D_{C/c(3)}\,,
\label{e:sigma_generic}
\end{align} 
with $f_{a/A(3)}$ the twist-3 non-perturbative function associated with parton $a$ in proton $A$, and likewise for the other distribution and fragmentation functions.
The hard factors corresponding to each term are given by $H$, $H'$, and $H''$, and the symbol $\otimes$ represents convolutions in the appropriate momentum fractions.  In Eq.~(\ref{e:sigma_generic}) a sum over partonic channels and parton flavors in each channel is understood.  One then defines 
\begin{equation}
A_N = \frac{\frac{1} {2} \left[d\sigma(\vec{S}_\perp) - d\sigma(-\vec{S}_\perp)\right]} {\frac{1} {2} \left[d\sigma(\vec{S}_\perp) + d\sigma(-\vec{S}_\perp)\right]} \equiv \frac{d\Delta\sigma(\vec{S}_\perp)} {d\sigma_{unp}}\,,
\end{equation}
where $d\sigma_{unp}$ is the unpolarized cross section.  Note of course that for $A_N^{jet}$ and $A_N^\gamma$ the third term in Eq.~(\ref{e:sigma_generic}) is not relevant, and the fragmentation functions $D_{C/c(2)}$ in the first two terms are not needed.\footnote{Here we ignore photons coming from fragmentation \cite{Gamberg:2012iq}, which can be largely suppressed by using isolation cuts.}

We see from Eq.~(\ref{e:sigma_generic}) that for the process of hadron production, which as we mentioned has been intensely studied for close to 40 years, the distribution and fragmentation twist-3 contributions to the cross section cannot be disentangled, i.e., all of them are summed together in the cross section.  This is different from reactions where transverse momentum dependent (TMD) functions are relevant, e.g., semi-inclusive deep-inelastic scattering (SIDIS) and electron-positron annihilation to two hadrons, where one can isolate certain effects~\cite{Boer:1997mf,Bacchetta:2006tn,Pitonyak:2013dsu}, e.g., Sivers~\cite{Sivers:1989cc} and Collins~\cite{Collins:1992kk} asymmetries.  Consequently, one must be careful in determining which term in Eq.~(\ref{e:sigma_generic}) is dominant.  For many years it was often assumed that the first term, specifically the piece involving the Qiu-Sterman function $G_F(x,x)$\footnote{There are different notations used in the literature for the Qiu-Sterman function, most notably $T_F(x,x)$.}~\cite{Qiu:1991pp,Qiu:1991wg,Qiu:1998ia}, was the main cause of $A_N^\pi$.  However, this led to a so-called ``sign mismatch'' between the Qiu-Sterman function and the TMD Sivers function extracted from SIDIS~\cite{Kang:2011hk}.  This issue could not be resolved through more flexible parametrizations of the Sivers function~\cite{Kang:2012xf} and, in fact, the authors of Ref.~\cite{Metz:2012ui} argued, by looking at $A_N$ data on the neutron target TSSA in inclusive DIS~\cite{Katich:2013atq}, that $G_F(x,x)$ cannot be the main source of $A_N^\pi$. Recently we showed in Ref.~\cite{Kanazawa:2014dca} for the first time that the fragmentation contribution in collinear twist-3 factorization (i.e., the third term in Eq.~(\ref{e:sigma_generic})) actually can describe $A_N^\pi$ very well.  By using a Sivers function fully consistent with SIDIS, we demonstrated that this mechanism could also resolve the sign-mismatch puzzle.  Nevertheless, an independent extraction of $G_F(x,x)$, through observables like $A_N^{jet}$ and $A_N^\gamma$, is crucial to confirm this assertion.  However, one must keep in mind that for $A_N^{jet}$ and $A_N^\gamma$ other twist-3 distribution effects can enter besides the Qiu-Sterman function.  Thus, in order to have a ``clean'' extraction of $G_F(x,x)$, it would be ideal if these other terms were numerically small. 

Therefore, in this paper we return to the study of the TSSA in $p^\uparrow p \to \gamma X$ to see if this reaction could provide such an observable.  (For recent analyses of this effect in $p^\uparrow A$ collisions, see Refs.~\cite{Kovchegov:2012ga,Schafer:2014zea}.)  In twist-3 collinear factorization, $A_N^\gamma$ has contributions from multiparton correlators inside either the transversely polarized proton (i.e., the first term in Eq.~(\ref{e:sigma_generic})) or the unpolarized proton (i.e., the second term in Eq.~(\ref{e:sigma_generic})).  For this process the former has been widely discussed in the literature for both (twist-3) quark-gluon-quark~\cite{Qiu:1991wg,Kouvaris:2006zy,Ji:2006vf,Koike:2006qv,Kanazawa:2011er,Gamberg:2012iq,Kanazawa:2012kt,Gamberg:2013kla} and tri-gluon~\cite{Koike:2011nx} non-perturbative functions.  For the former one has both soft-gluon pole (SGP)  effects (i.e., the Qiu-Sterman function) as well as soft-fermion pole (SFP) terms. We focus here instead on the quark-gluon-quark correlator that exists inside the unpolarized proton, where again both SGPs and SFPs enter. (Note that tri-gluon correlators only arise inside a transversely polarized hadron.)  This twist-3 function $E_F(x_1,x_2)$, defined below, is a chiral-odd object that combines with the twist-2 chiral-odd transversity to bring about the asymmetry. Its impact on TSSAs so far has been investigated only for light-hadron production~\cite{Kanazawa:2000hz}, and its contribution to $A_N^\pi$ was found to be negligible~\cite{Kanazawa:2000kp}.  However, its significance for $A_N$ in other processes is still unclear. 
In fact, the effect from $E_F(x_1,x_2)$ could be enhanced for
$A_N^\gamma$ compared with $A_N^\pi$ just like that
from $G_F(x_1,x_2)$ is significantly enhanced because of the color
structure
\cite{Kang:2011hk,Kanazawa:2012kt}.  In addition, we summarize the previous results for $A_N^\gamma$ that enter from the first term in Eq.~(\ref{e:sigma_generic}) and perform a numerical study at RHIC kinematics of all quark-gluon-quark correlator contributions. Again, this complete solution includes pieces involving chiral-even and chiral-odd functions at both SGPs and SFPs.  We note that previous work on tri-gluon correlators show such contributions for this observable are negligible in the forward region~\cite{Koike:2011nx}.  We find then through our results that $A_N^\gamma$ could give a ``clean'' observable to extract the Qiu-Sterman function.  Since we use the Sivers function taken from SIDIS to develop our input for the Qiu-Sterman function, one can also learn about the process dependence of the Sivers function from such experiments.  Furthermore, by comparing our results to those that already exist in the literature from the Generalized Parton Model (GPM)~\cite{D'Alesio:2004up, Anselmino:2013rya}, we find that this observable could differentiate between the GPM and twist-3 formalisms.

The rest of the paper is organized as follows: in Sec.~\ref{s:calc} we provide calculational details on the second term in Eq.~(\ref{e:sigma_generic}) for $A_N^\gamma$; in Sec.~\ref{s:phen} we report on our phenomenological analysis, including error bands generated from the uncertainties of our non-perturbative inputs, in particular the Sivers function; and in Sec.~\ref{s:sum} we summarize and conclude our work.  Some technical steps of the calculation are presented in Appendix A.

%
%
\section{Calculation of the unpolarized quark-gluon-quark correlator term} \label{s:calc}

In this section we provide the main steps of the analytical calculation of the contribution to $A_N^\gamma$ that arises from twist-3 effects in the unpolarized proton.  First, we define the relevant functions needed in our computation.  The unpolarized quark-gluon-quark correlator for a quark of flavor $a$ is defined
as~\cite{Jaffe:1991ra,Boer:1997bw,Ma:2003ut,Zhou:2009jm}\footnote{Eq.~(\ref{e:E_F}) holds in the lightcone gauge $A^+ = 0$, where gauge links between the field operators reduce to unity.}
\begin{align}
M_{F, ij}^{a,\alpha} (x_1,x_2)&\equiv\int \!\frac{d\lambda}{2\pi} \frac{d\mu}{2\pi}
 e^{i\lambda x_1} e^{i\mu (x_2-x_1)} \bra{p} \psib_j^a (0) gF^{\alpha
 n}(\mu n) \psi_i^a (\lambda n) \ket{p} \nn \\
 &= \frac{M_N}{4} ( \gamma_5 \Slash{p} \gamma_\lambda )_{ij}
  \epsilon^{\lambda\alpha np} E_{F}^a(x_1,x_2)+\cdots\,, \label{e:E_F}
\end{align}
where $M_N$ is the nucleon mass, $F^{\mu\nu} \equiv F^{\mu\nu}_A T_A$ is the field strength tensor with $T_A$ being the color matrices, $g$ is the strong coupling constant, $\epsilon^{\lambda\alpha np}=\epsilon^{\lambda\alpha \rho\sigma}
n_{\rho}p_{\sigma}$ with $\epsilon_{0123} = +1$ is the Levi-Civita tensor,
and $n^\mu$ is a light-like vector satisfying $p\cdot n=1$ and $n\cdot\Sp=0$.  The dots represent higher-twist contributions.
From Hermiticity and time-reversal invariance, this function has
the symmetry property 
\begin{equation}
 E_{F}^a(x_1,x_2) = E_{F}^a(x_2,x_1)\,.
\end{equation}
The antiquark distribution is given by 
\begin{equation}
 E_{F}^{\bar{a}}(x_1,x_2) = E_{F}^a (-x_2,-x_1)\,.
\end{equation}
The SGP function $E_F^a(x,x)$ is related to the Boer-Mulders
function $h_1^{\perp a}(x,\vec{k}_T^2)$, defined through an unpolarized quark-quark correlator as~\cite{Boer:1997nt,Goeke:2005hb}
\begin{equation}
 \int\!\frac{d y^- d^2\vec{y}_T} {(2\pi)^3}e^{ik\cdot y}\langle p|\bar{\psi}^a_j(0)\mathcal{W}_{\pm}(0,y)\psi^a_i(y)|p\rangle \bigg |_{y^+ = \,0} = \frac{1}{2M_N} \sigma^{\alpha\nu} k_{\perp\alpha}\,p_\nu\,h_1^{\perp a\,(\pm)}(x,\vec{k}_T^2) + \dots\,,
\end{equation}
where $\mathcal{W}_{\pm}(0,y)$ is a future-pointing (+) or past-pointing ($-$) Wilson line connecting the quark fields between their spacetime points, which renders the bilocal matrix element color gauge invariant.  Which Wilson line is chosen depends on the process under consideration --- see, e.g.,~\cite{Collins:2002kn, Ji:2002aa, Belitsky:2002sm, Boer:2003cm, Bomhof:2004aw, Bacchetta:2005rm}.  Note that since $h_1^\perp$ is na\"{i}ve time-reversal odd, one has $h_1^{\perp (+)} = -h_1^{\perp (-)}$~\cite{Collins:2002kn}.  Using
$\gamma_5\Slash{p}\gamma_\lambda \epsilon^{\lambda\alpha
np}=\sigma^{\alpha p}$ for $\alpha=\perp$ leads to the relation~\cite{Boer:2003cm}
\begin{equation}
 E_{F}^a(x,x) = \pm \frac{1}{\pi M_N^2} \int \!d^2\vkT \,\vkT^2 \,h_1^{\perp a\, (\pm)}\label{e:EFBM}
  (x, \vkT^2)\,.
\end{equation}
The quark transversity $h(x)$ is given by a polarized quark-quark correlator in the standard way,
\begin{equation}
 \int\! \frac{d\lambda}{2\pi} e^{i\lambda x} \bra{p\Sp} \psib_j^a (0)
  \psi_i^a (\lambda n) \ket{p\Sp} = \frac{1}{2} (\gamma_5 \Slash{S}_\perp
  \Slash{p})_{ij} h^a(x)+\cdots\,.
\end{equation}

\begin{figure} 
 \begin{center}
 \fig{0.6}{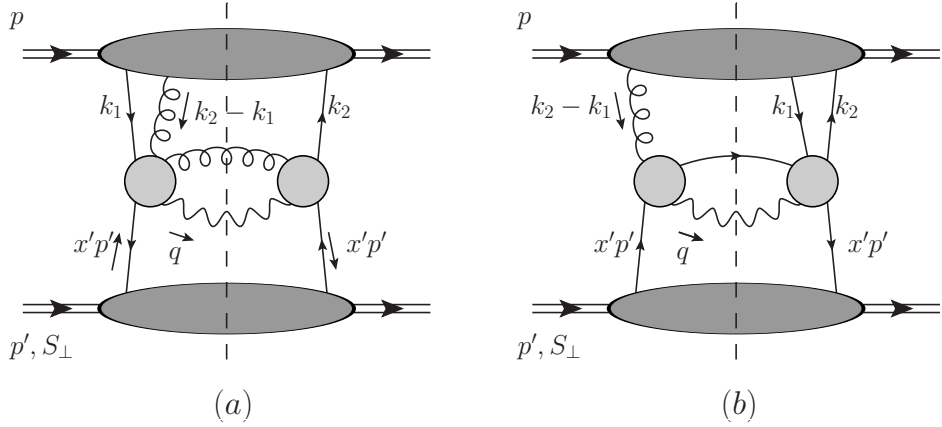}
 \caption{Generic diagrams giving rise to $A_N^\gamma$ from the chiral-odd twist-3
  function $E_F(x_1,x_2)$ in the unpolarized proton are shown in (a), (b). The mirror diagrams also contribute to
  the asymmetry.} \label{f:gen_dia}
 \end{center}
\end{figure} 

For direct photon production in proton-proton collisions,
\begin{eqnarray}
 \pup (p',\Sp) + p (p) \to \gamma (q) + X\,,
\end{eqnarray}
we define the Mandelstam variables at the hadron level by
\begin{eqnarray}
 S = (p'+p)^2,\quad T = (p'-q)^2,\quad  U = (p-q)^2,
\end{eqnarray}
and at the partonic level by
\begin{eqnarray}
 \sh = (x'p' + xp)^2, \quad \th = (x'p' - q)^2, \quad \uh = (xp - q)^2.
\end{eqnarray}
Generic diagrams for the contribution to $A_N^\gamma$ we wish to compute are shown in Fig.~\ref{f:gen_dia}.  We will denote this as the ``chiral-odd'' ($\chi$-$o$) piece since it involves chiral-odd functions.

\begin{figure} 
 \begin{center}
  \fig{0.6}{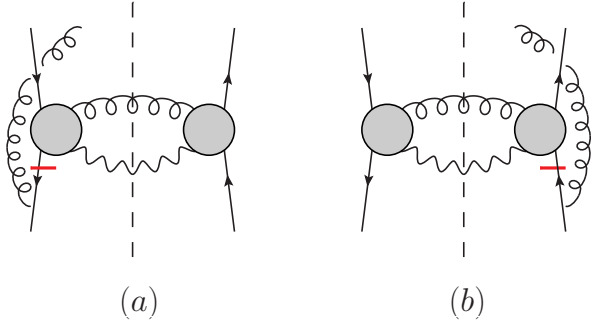}
  \caption{Feynman diagrams that produce SGPs are shown in (a), (b). The circle
  represents the sum of the $t$-channel and $u$-channel diagrams for the
  partonic subprocess. \label{SGP}}
 \end{center}
\end{figure}

\begin{figure}
 \begin{center}
  \fig{0.4}{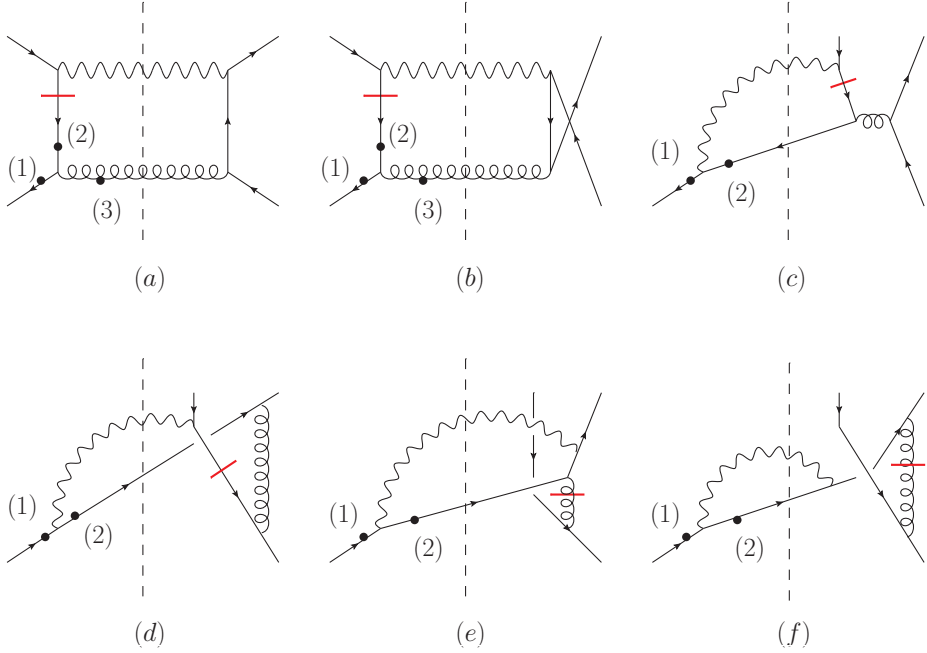}
  \caption{Feynman diagrams that produce SFPs are shown in (a)--(f) . A coherent
  gluon line from the unpolarized proton attaches to each dot. The
  mirror diagrams also contribute. Graphs producing SFPs 
that cancel one another are not shown. See \cite{Kanazawa:2011er} for details. \label{SFP}}
 \end{center}
\end{figure}

Based on the collinear twist-3 formalism in the Feynman
gauge for the pole contributions 
\cite{Eguchi:2006mc,Beppu:2010qn}, it is straightforward to obtain
the cross section in a gauge invariant
factorized form as 
\begin{equation}
E_\gamma\frac{d^3\Delta\sigma^{\chi\textit{-}o}}{d^3\vec{q}} =
  \frac{e_a^2 \alpha_{em}\alpha_s}{S} \int\! \frac{dx'}{x'} h(x') \int\!
  dx_1\!\int\! dx_2 \,\Tr{iM_F^\alpha (x_1,x_2) \del{S_\sigma (k_1,k_2,x'p',q)
  p^\sigma}{k_2^\alpha} \cl} ,
\end{equation}
where $\alpha_{em(s)} = e^2(g^2)/4\pi$ is the electromagnetic (strong)
coupling constant, $S_\sigma (k_1,k_2,x'p',q)$ is the hard factor
corresponding to the quark-gluon-quark correlator in the unpolarized
proton $\langle p | \bar{\psi} A^\sigma \psi |p\rangle$, and
``c.l.'' indicates the collinear limits: 
$k_{1(2)}\to x_{1(2)}p$. For
simplicity, we omitted the color and flavor indices, $A$ and $a$, in
$M^\alpha_F(x_1,x_2)$ and $S_\sigma(k_1,k_2,x'p',q)$, and the summation over these indices is understood. 
We recall that for the calculation of the SFP term one can
make a simplification owing to the Ward identity for the hard part \cite{Eguchi:2006mc}, 
\begin{eqnarray}
 \del{S_\sigma (k_1,k_2,x'p',q) p^\sigma}{k_2^\alpha} \cl^{\rm SFP} = \frac{1}{x_1-x_2} S_\alpha^{\rm SFP}(x_1p,x_2p,x'p',q)\,.
\end{eqnarray}
With the factorized expression at hand, we now turn to calculate
the perturbative hard coefficients. The relevant diagrams are shown in
Figs.\,\ref{SGP} and \ref{SFP} for the SGP and SFP contributions, respectively.   
For the calculation of the former
it is convenient to utilize the master
formula \cite{Koike:2006qv,Koike:2007rq,Koike:2011ns}, which we derive in Appendix A
for the present case. 
Recognizing that the SGP occurs through the initial-state interaction, the corresponding
hard part 
can be reduced as
\begin{equation}
 \del{S_\sigma(k_1,k_2,x'p',q)p^\sigma}{k_2^\alpha} \cl^{\rm SGP} =
  -i\pi\delta(x_2-x_1) \left[ \frac{d}{d(x'p')^\alpha} S(x_1p,x'p',q) -
			\St_\alpha (x_1p,x'p',q) \right],  \label{relation}
\end{equation}
where $S(xp,x'p',q)$ represents a reduced cut 2-to-2 amplitude which becomes 
the corresponding cross section after
the Dirac- and color- traces are taken together with the
appropriate factor from $M^\alpha_F(x_1,x_2)$.  
Its precise definition is given in Appendix~A.
\hspace{-0.15cm}Here we note the appearance of an additional term $\St_\alpha$,
which is absent for the SGP contributions involving chiral-even functions.
It is given by 
\begin{equation}
 \St_\alpha (xp,x'p',q) = S(xp,x'p',q) \vb_{ \Slash{S}_\perp \to \frac{
  S_{\perp\alpha} \Slash{p}}{x' p\cdot p'} }\,.
\end{equation}
The SGP hard cross section for the first term in (\ref{relation}) is
closely related to the
transverse double-spin dependent hard part for the leading-twist
observable $A_{TT}$ in direct photon production~\cite{Mukherjee:2003pf}, 
\begin{equation}
 \sigmah_{TT} = \frac{4C_F}{N} \frac{\sh}{\th\uh} 
\left[ 2 (q\cdot\Sp) (q\cdot\Sp') + \frac{\th\uh}{\sh} (\Sp\cdot\Sp') \right].
\end{equation}
The final result is given by
\begin{align} \label{e:co_result}
 E_\gamma\frac{d^3\Delta\sigma^{\chi\textit{-}o}}{d^3\vec{q}} &=
  \frac{\alpha_{em}\alpha_s \pi M_N}{S} \epsilon^{pnq\Sp} \int
  \!\frac{dx'}{x'} \int\! \frac{dx}{x}\, \delta(\sh+\th+\uh) \nn\\
 & \hspace{0.5cm}\times \sum_a e_a^2 \left[\left( E_F^a(x,x) - x \frac{dE_F^a(x,x)}{dx} \right)
       h^{\bar{a}} (x')
  \frac{\sigmah_1^{\rm SGP}}{-\th} 
   + E_F^a(x,x) h^{\bar{a}} (x')
  \sigmah_2^{\rm SGP} \right],
\end{align}
where the partonic hard parts read\footnote{Since we have already factored out $\epsilon^{pnq\Sp}$
in (\ref{e:co_result}), we must divide by this factor after the first equality in (\ref{e:co_sigma1}).}
\begin{align}
 \sigmah^{\rm SGP}_1 &=
  \frac{1} {\epsilon^{pnq\Sp}}\,\sigmah_{TT}(\Sp^\lambda\to\epsilon^{pnq\lambda},\Sp'\to\Sp, C_F/N\to
 -1/2N^2)= -\frac{2}{N^2}\,, \label{e:co_sigma1} \\
 \sigmah^{\rm SGP}_2 &= \frac{(\th-\uh)}{N^2\th\uh}\,,\label{e:co_sigma2}
\end{align}
with $N=3$ the number of colors and $C_F = 4/3$.  It turns out
that the hard factors for the SFP pieces vanish after summing over all graphs.  

%
%
\section{Phenomenology of $A_N^\gamma$} \label{s:phen}

In this section we summarize the other pieces needed, in addition to Eq.~(\ref{e:co_result}), in order to give a numerical estimate of $A_N^\gamma$ from quark-gluon-quark correlators in the proton.  First, the unpolarized cross section is given by
\begin{equation}
 E_\gamma \frac{d^3\sigma_{unp}}{d^3\vec{q}} = \frac{\alpha_{em} \alpha_s}{S}\frac{1} {N}\! \int
 \! \frac{dx'}{x'} \int \!\frac{dx}{x}\,\delta (\sh+\th+\uh)  \sum_a e_a^2 
  \Big[ f^a (x) f^{\bar{a}}(x') \sigmah_{a\bar{a}} + f^a (x) f^g(x')
   \sigmah_{ag} + f^g(x) f^{a} (x') \sigmah_{ga} \Big],
\end{equation}
with the hard cross sections
\begin{align}
 \sigmah_{a\bar{a}} &= 2C_F \left( \frac{\uh}{\th} + \frac{\th}{\uh} 
			     \right) = \hat{\sigma}_{\bar{a}a}\,,\label{e:unp_aabar} \\
 \sigmah_{ag} &= 2 T_R \left( \frac{\sh}{-\uh} + \frac{-\uh}{\sh} \right), \\
 \sigmah_{ga} &= 2 T_R \left( \frac{\sh}{-\th} + \frac{-\th}{\sh}
			\right)\label{e:unp_ga}\,,
\end{align}
where $T_R = 1/2$.  The functions $f^a(x)$ and $f^g(x)$ are the usual unpolarized quark and gluon distribution functions, respectively.

Next, we have the spin-dependent cross section involving chiral-even functions, which we will call the ``chiral-even'' ($\chi$-$e$) piece.  These terms involve the functions $G_F(x_1,x_2)$ and $\Gt_F(x_1,x_2)$.  We refer to Ref.~\cite{Eguchi:2006qz} for the definitions and symmetry properties of these correlators.  As mentioned in Sec.~\ref{s:intro}, $G_F(x,x)$ is known as the Qiu-Sterman function and is related to the Sivers function~\cite{Sivers:1989cc} $f_{1T}^\perp(x,\vec{k}_T^2)$ through~\cite{Boer:2003cm} 
\begin{equation}
 G_{F}^a(x,x) = \mp \frac{1}{\pi M_N^2} \int \!d^2\vkT \,\vkT^2 \,f_{1T}^{\perp a\, (\pm)}\label{e:GFSiv}
  (x, \vkT^2)\,.
\end{equation}
Recall also from Sec.~\ref{s:intro} the sign mismatch between two different extractions of the Qiu-Sterman function~\cite{Kang:2011hk}.  This seems to have a potential resolution by attributing $A_N^\pi$ to the fragmentation mechanism rather than the Qiu-Sterman function~\cite{Kanazawa:2014dca}, but still an independent extraction of $G_F(x,x)$ through an observable like $A_N^\gamma$ is necessary.  As with the chiral-odd term, the chiral-even cross section has both SGPs and SFPs.  These are given, respectively, by~\cite{Qiu:1991wg,Kouvaris:2006zy,Ji:2006vf,Koike:2006qv,Kanazawa:2011er,Gamberg:2012iq,Kanazawa:2012kt,Gamberg:2013kla} 
\begin{align}
E_\gamma \frac{d^3\Delta\sigma^{\chi{\textit -}e,\rm SGP}}{d^3\vec{q}} &= -\frac{\alpha_{em} \alpha_s}{S} \frac{\pi M_N} {NC_F} 
 \epsilon^{pnq\Sp} \int\!
  \frac{dx'}{x'} \int\!\frac{dx}{x}\, \delta (\sh+\th+\uh) \nn\\
 &\hspace{0.3cm} \times \sum_a e_a^2 \frac{1}{-\uh}
  \left[ \frac{1}{2N} f^{\bar{a}}(x) \sigmah_{\bar{a}a} - \frac{N}{2}
  f^g(x) \sigmah_{ga} \right] \left[ x'\,\frac{dG_F^a(x',x')}{dx'} - G_F^a(x',x') \right], \label{e:ceSGP}\\
E_\gamma \frac{d^3 \Delta\sigma^{\chi{\textit -}e,\rm SFP}}{d^3 \vec{q}} &=
-\frac{\alpha_{em}\alpha_s}{S} \frac{\pi M_N}{2N} \eps^{pnq\Sp} \int\!
\frac{dx'}{x'} \int\! \frac{dx}{x}\, \delta (\sh+\th+\uh) \nn\\
&\hspace{0.3cm}\times \sum_a
\bigg[ \sum_{b} e_a e_b \sigmah^{\rm SFP}_{ab} \left\{ G_F^a
(0,x') + \Gt_F^a (0,x') \right\} f^b (x)  \nn\\
&\hspace{0.3cm} \quad + \sum_{b} e_a e_b \sigmah^{\rm SFP}_{a\bar{b}}\left\{ G_F^a
(0,x') + \Gt_F^a (0,x') \right\} f^{\bar{b}} (x) \nn\\
&\hspace{0.3cm} \qquad +\,  e_a^2 \sigmah^{\rm SFP}_{ag}\left\{ G_F^a
(0,x') + \Gt_F^a (0,x') \right\} f^g(x) \bigg] , \label{e:ceSFP}
\end{align}
where $\sigmah_{\bar{a}a}$, $\sigmah_{ga}$ in the SGP term are given in (\ref{e:unp_aabar}), (\ref{e:unp_ga}).  Those for the SFP contribution 
read \cite{Kanazawa:2011er}
\begin{align}
 \sigmah_{ab}^{\rm SFP} &= \frac{2(\sh^2+\uh^2)}{\th^2\uh} 
+ \frac{2\sh(\uh-\sh)}{N\th\uh^2} \delta_{ab}\,,\\
\sigmah_{a\bar{b}}^{\rm SFP} &= -\frac{2(\sh^2+\uh^2)} {\th^2\uh}
+\left[ \frac{2N\sh}{\uh^2}+\frac{2(\uh^2+\sh\th)}{N\sh\th\uh}\right]\delta_{ab}\, ,\\
\sigmah_{ag}^{\rm SFP} &= \frac{2[N^2\th\uh-\sh(\sh-\th)]}{(N^2-1)\sh\th\uh}\,.  
\end{align}
The sum for $a$ is over all quark and antiquark flavors ($a = u, \,d,\, s,\, \bar{u},\, \bar{d},\, \bar{s}$), and $\sum_b$ indicates that the sum for $b$ is restricted over the quark flavors when $a$ is a quark and over antiquark flavors when $a$ is an antiquark.

Finally, we are in a position to give a numerical estimate for $A_N^\gamma$ from the sum of Eqs.~(\ref{e:co_result}), (\ref{e:ceSGP}), (\ref{e:ceSFP}).  This requires inputs for the non-perturbative functions that enter the formulas.  For the SGP correlators $E_F(x,x)$ and $G_F(x,x)$ we make use of the identities in Eqs.~(\ref{e:EFBM}), (\ref{e:GFSiv}), respectively, that relate the first to the Boer-Mulders function and the second to the Sivers function.  For the Boer-Mulders function we take the parametrization from Ref.~\cite{Barone:2010gk} (see \cite{Barone:2009hw} for a fit without antiquarks), while for the Sivers function we use the extraction from Ref.~\cite{Anselmino:2008sga}.  Since at
this point no information on the SFP functions is available, we assume
the relation~\cite{Koike:2009ge}
\begin{align}
G_F(0,x)+\Gt_F&(0,x) = G_F(x,x)\,. \label{e:SFP_func}
\end{align}
We mention that it will be necessary to obtain these SFP functions through other reactions in order to give a firm prediction of the effect these have on $A_N^\gamma$.  Nevertheless, model calculations of quark-gluon-quark correlators show that chiral-even SFP functions are much smaller as compared to the SGP one~\cite{Braun:2011aw} and might even vanish~\cite{Kang:2010hg}.  Therefore, we believe that the ansatz in (\ref{e:SFP_func}) is suitable to obtain the maximum possible chiral-even SFP contribution to $A_N^\gamma$.  The transversity function is taken from the extraction in~\cite{Anselmino:2013vqa}, with the antiquarks set to saturate the Soffer bound~\cite{Soffer:1994ww}.  For the unpolarized distributions we use the GRV98~\cite{Gluck:1998xa} fit, which were also used by Refs.~\cite{Barone:2009hw,Barone:2010gk,Anselmino:2008sga,Anselmino:2013vqa} for the other aforementioned non-perturbative inputs that enter.  All parton correlation functions are evaluated at the scale $q_T$ with leading order evolution of the collinear functions.

\begin{figure}
 \begin{center}
  \fig{0.6}{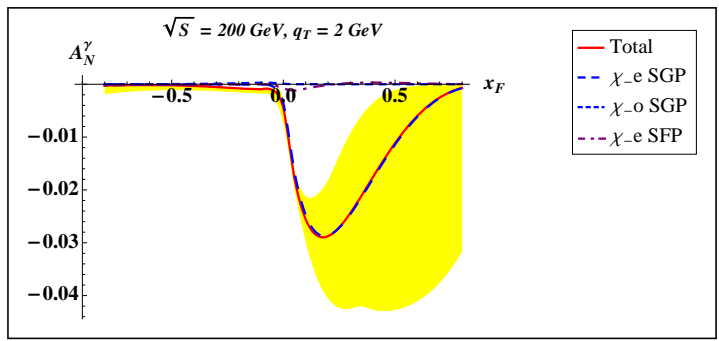}
 \end{center}
 \vspace{-0.65cm}
 \caption{$A_N^\gamma$ vs.~$x_F$ at a fixed $q_T = 2\,{\rm GeV}$ for a range $|x_F| < 0.8$.  The center-of-mass energy is set at  $\sqrt{S} = 200\,{\rm GeV}$. The SGP pieces are given by the long-dashed curve for chiral-even and the short-dashed curve for chiral-odd.  The chiral-even SFP part is shown by the dot-dashed curve.  Note again that the chiral-odd SFP term vanishes.  The sum of all contributions is the solid curve.  The shaded area gives the error band in the calculation as described in the text. \label{pt2}}
\end{figure}

\begin{figure}
 \begin{center}
  \fig{0.5}{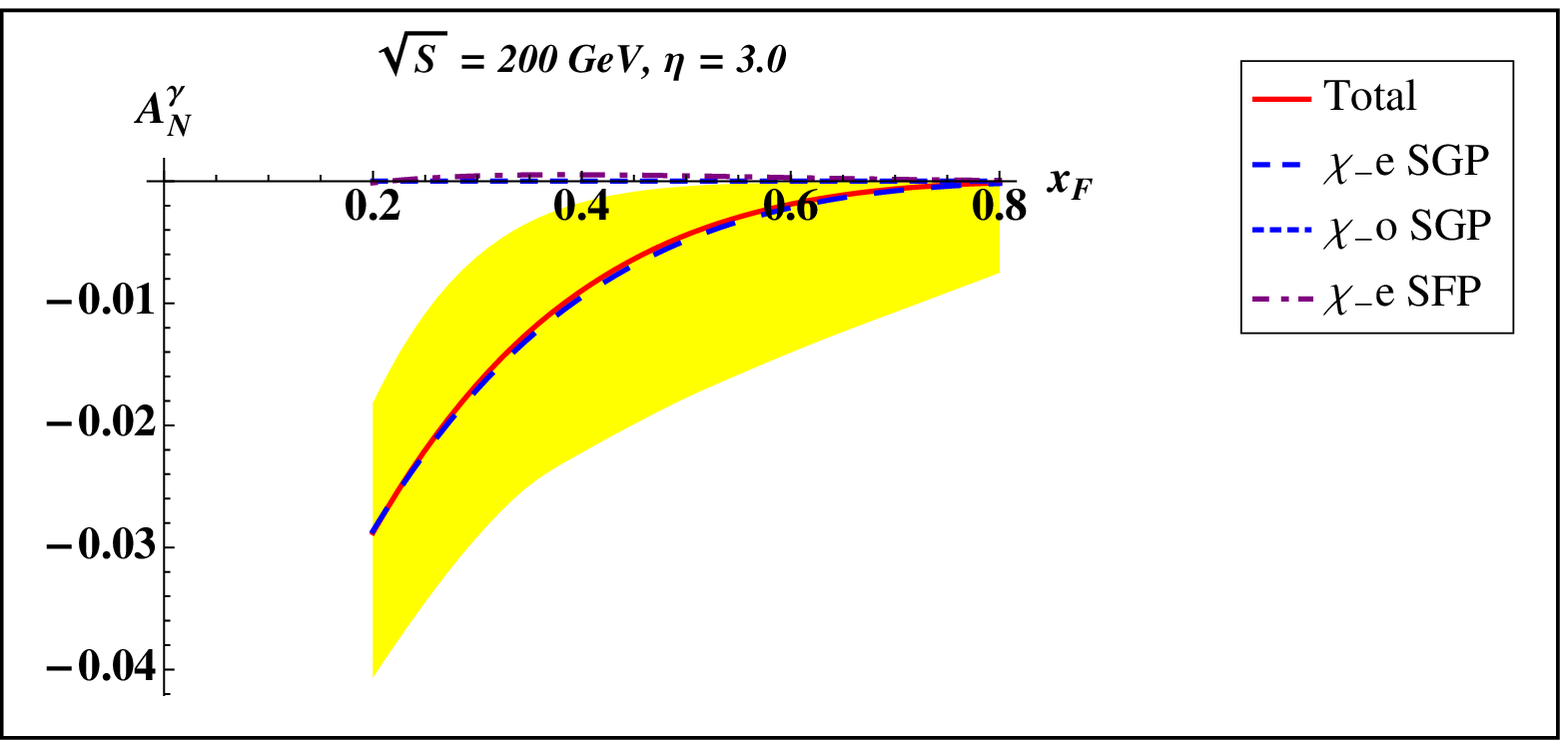}
  \fig{0.495}{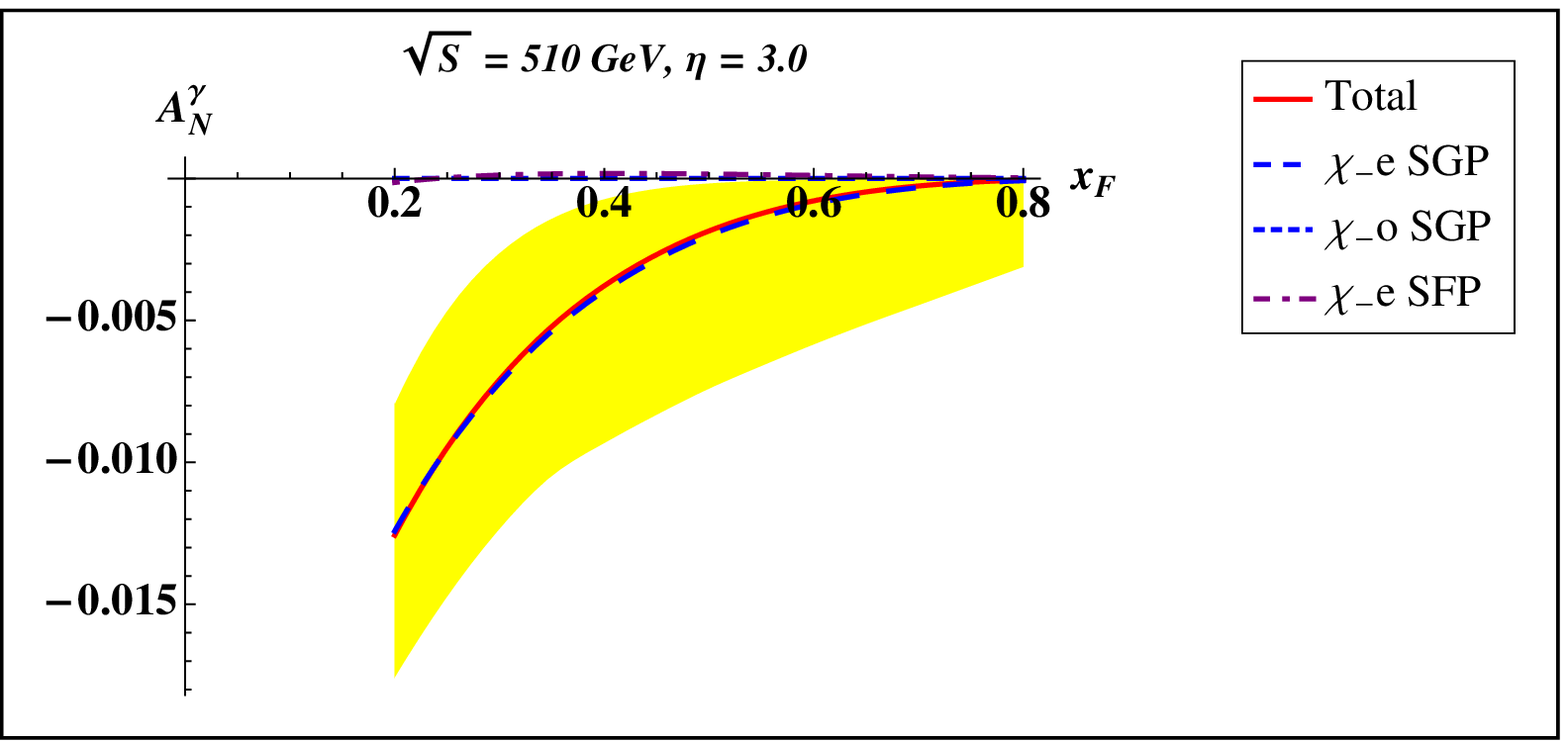}
  \fig{0.499}{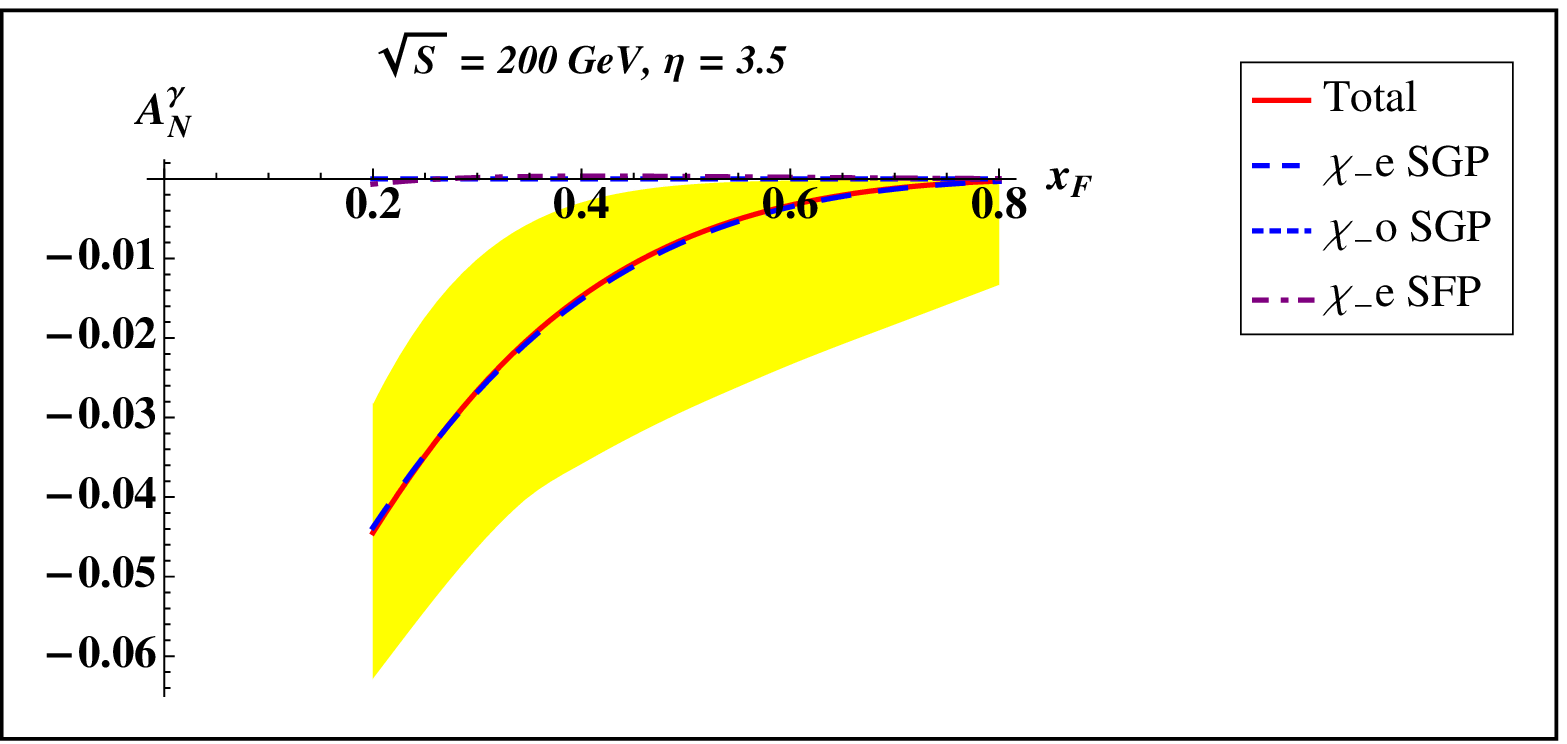} 
  \fig{0.497}{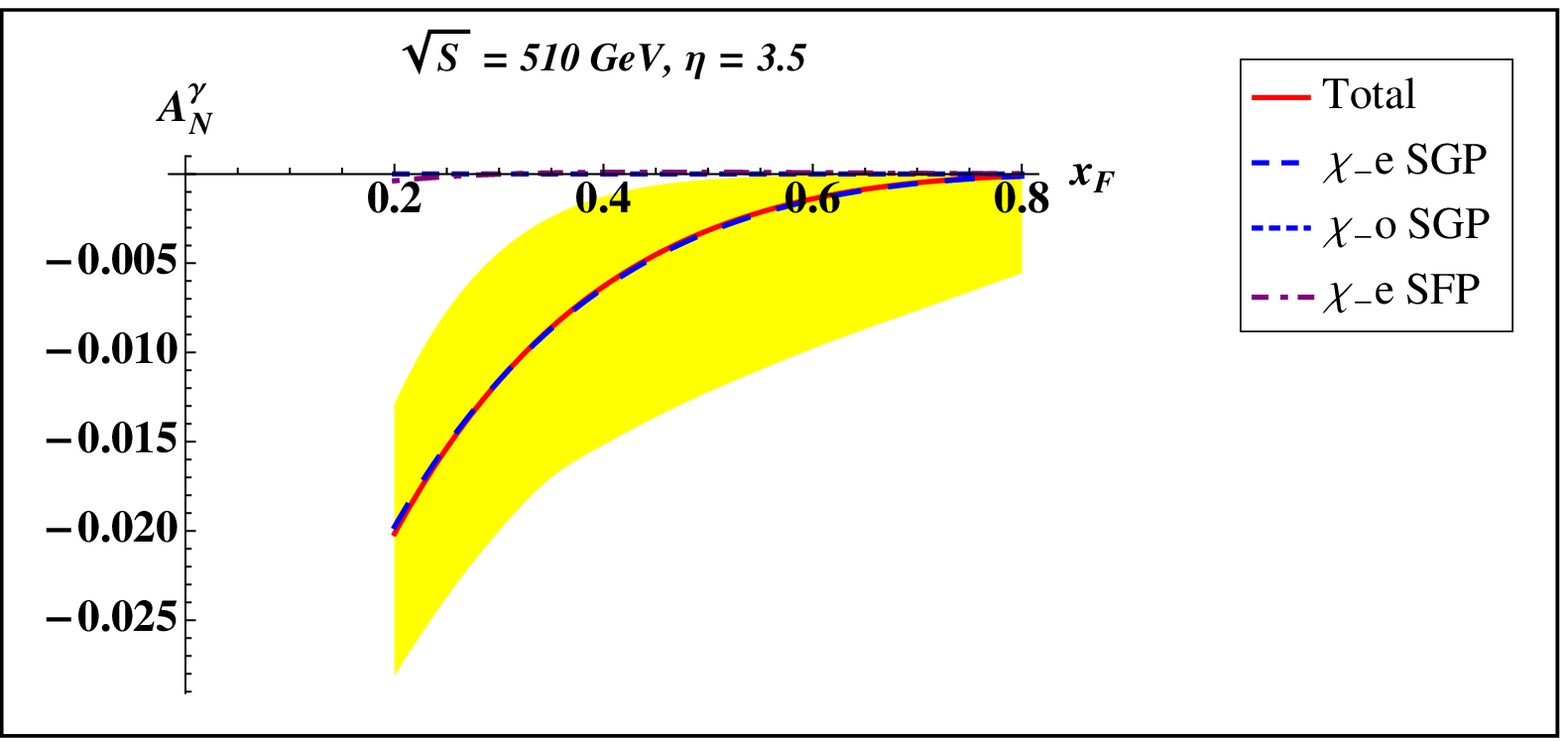}
 \end{center}
 \vspace{-0.65cm}
 \caption{$A_N^\gamma$ vs.~$x_F$ at fixed $\eta$ for $\eta = 3.0,\,3.5$ and $\sqrt{S} = 200,\,510\,{\rm GeV}$.  The curve labels are the same as in Fig.~\ref{pt2}. \label{fixeta}}
\end{figure}

\begin{figure}
 \begin{center}
  \fig{0.5}{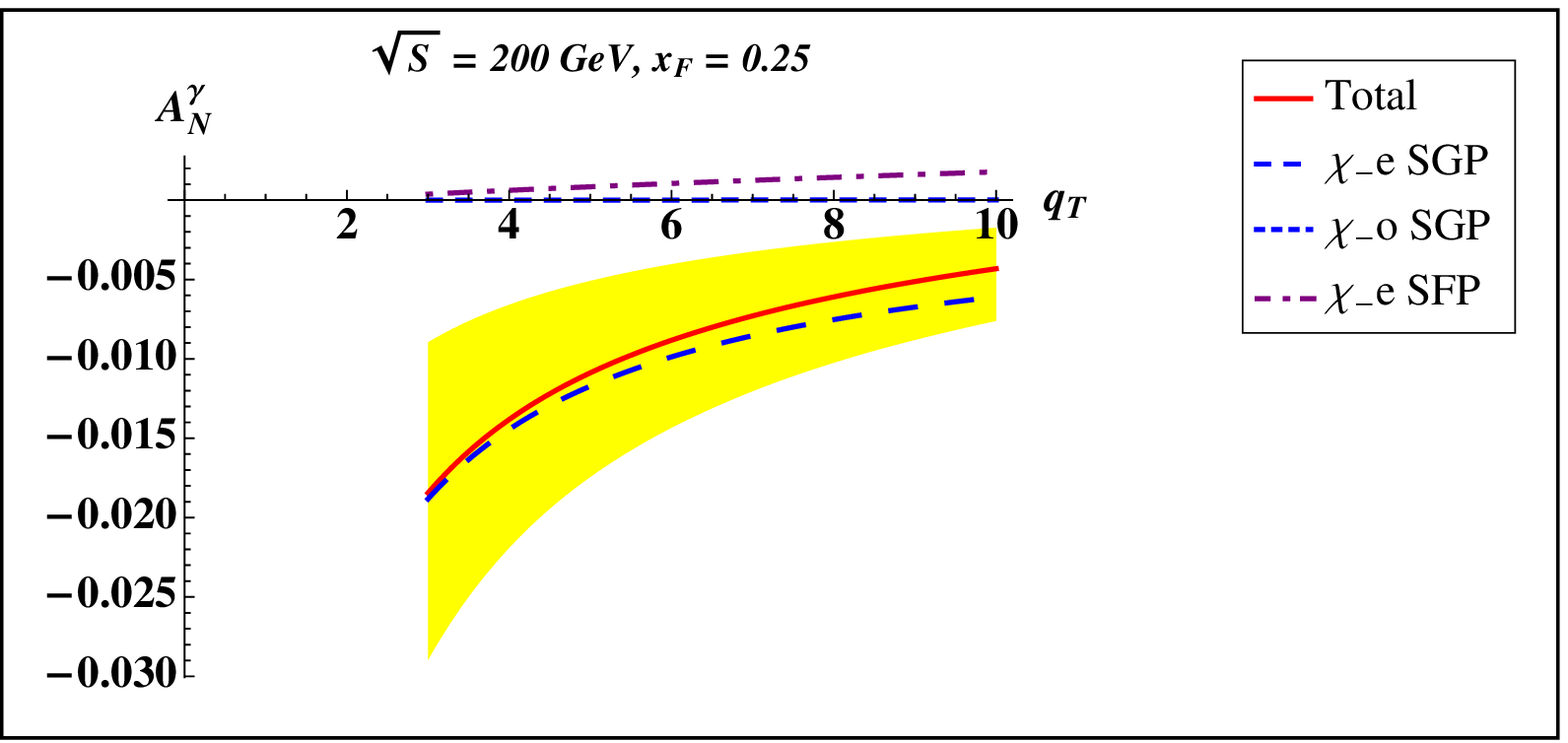}
  \fig{0.505}{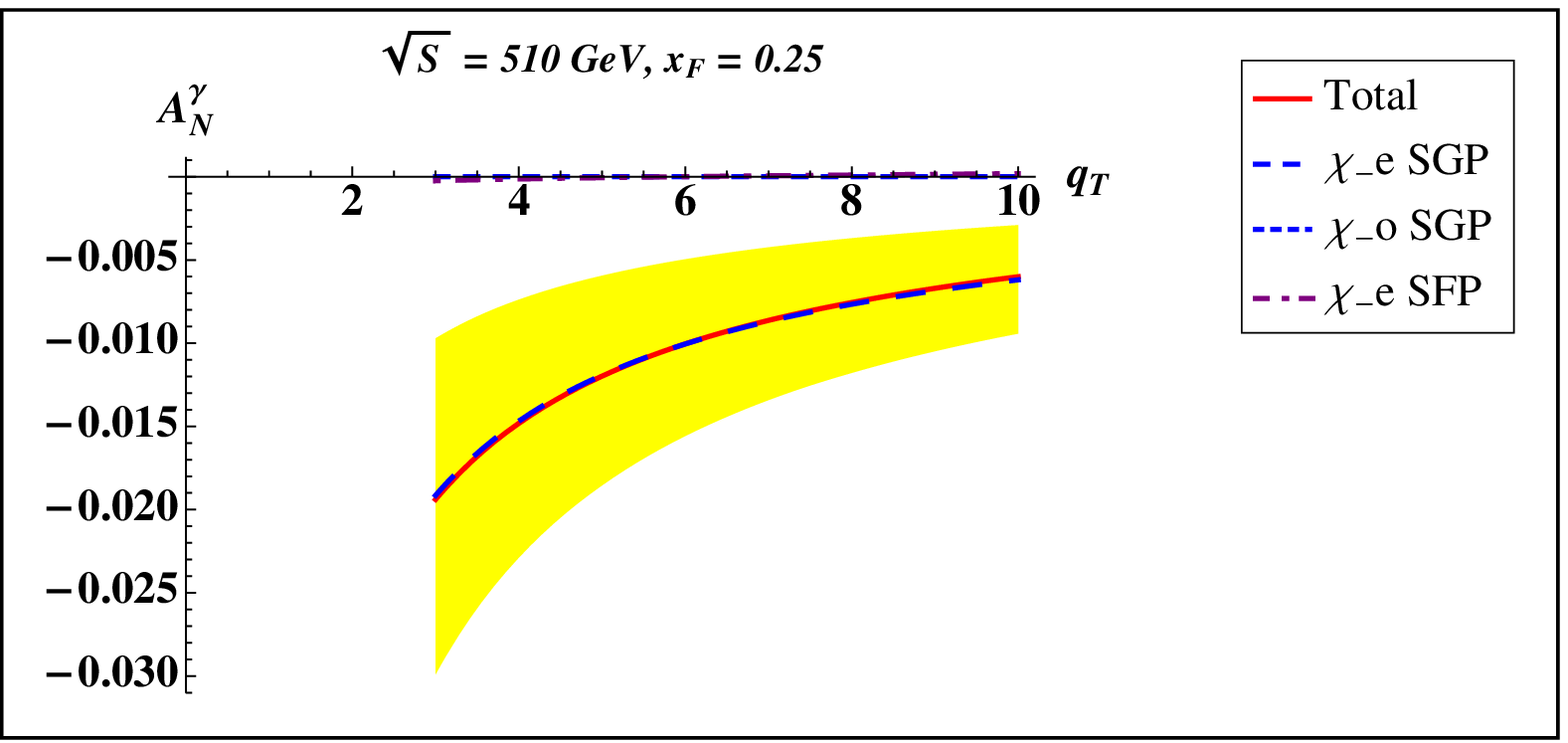}
 \end{center}
 \vspace{-0.65cm}
 \caption{$A_N^\gamma$ vs.~$q_T$ at a fixed $x_F = 0.25$ for $\sqrt{S} = 200,\,510\,{\rm GeV}$.  The curve labels are the same as in Fig.~\ref{pt2}. \label{fixxf}}
\end{figure}

In Figs.~\ref{pt2}--\ref{fixxf} we show an estimate of $A_N^\gamma$ from
all these pole contributions at different kinematics relevant for the
measurement of this observable at PHENIX and STAR.  As seen in
Figs.~\ref{pt2}--\ref{fixxf}, in general the chiral-odd piece is
negligible.  We have checked that this is a robust statement by changing
parameters in both the Boer-Mulders function from~\cite{Barone:2010gk}
(including looking at both Fit 1 and Fit 2) and the transversity
function from~\cite{Anselmino:2013vqa} within their error ranges.  In all cases, we found the chiral-odd piece is roughly four orders of magnitude smaller than the total asymmetry (in the $x_F$-range where $A_N^\gamma$ is nonzero).  In particular, this observation does not change if one allows for a flavor-dependent large-$x$ behavior in the Boer-Mulders or transversity functions, as the corresponding partonic hard cross sections (\ref{e:co_sigma1}), (\ref{e:co_sigma2}) are extremely small.  Our focus then is on the chiral-even contribution in Eqs.~(\ref{e:ceSGP}), (\ref{e:ceSFP}).  Since both the SGP and SFP terms are parameterized in terms of the Sivers function, we not only give central curves for these parts but also provide an error estimate for $A_N^\gamma$ based on the uncertainty in the Sivers function computed in~\cite{Anselmino:2008sga}.  That is, the band in the figures is the total error for $A_N^\gamma$ coming from Eqs.~(\ref{e:ceSGP}), (\ref{e:ceSFP}).  This uncertainty is quite large in the forward region due to the fact that the Sivers function is mostly unconstrained at large $x$~\cite{Anselmino:2008sga}.  

We now specifically discuss each figure.  In Fig.~\ref{pt2} we give $A_N^\gamma$ vs.~$x_F$ at a fixed $q_T = 2\,{\rm GeV}$ for a range $|x_F| < 0.8$. The center-of-mass energy is set at  $\sqrt{S} = 200\,{\rm GeV}$.  We see in the forward region that the asymmetry can be on the order of $\sim \!-(2-3\%)$, while in the backward region the effect is extremely small.  The ``Sivers-type'' $G_F(x,x)$ term essentially gives all of $A_N^\gamma$ at $x_F > 0$.  Recall from Ref.~\cite{Koike:2011nx} that the contribution from tri-gluon correlators to $A_N^\gamma$ was negligible for $x_F > 0$ but potentially significant for $x_F < 0$.  Thus, a measurement of $A_N^\gamma$ in the backward region could provide a constraint on these mostly unknown tri-gluon twist-3 functions.

In Fig.~\ref{fixeta} we give $A_N^\gamma$ vs.~$x_F$ at a fixed rapidity $\eta$ for $\eta = 3.0,\,3.5$ and $\sqrt{S} = 200,\,510\,{\rm GeV}$.  All of the plots present the same feature, namely, $A_N^\gamma$ decreases with increasing $x_F$.  The 200 GeV curves show an asymmetry of $\sim \!-(3-5\%)$ for $0.2<x_F<0.4$, while the 510 GeV ones give $A_N^\gamma\sim \!-(1-2\%)$ in that $x_F$-range.  The effect also increases with increasing rapidity.  Again, the Qiu-Sterman piece gives the entire asymmetry.\footnote{We found that the chiral-even SFP term can only become comparable to the SGP term if its associated non-perturbative functions fall off as slow or slower than the {\it derivative} of the Qiu-Sterman function at large $x$.  This is an unlikely scenario, as we stated after Eq.~(\ref{e:SFP_func}).}  We note that in Ref.~\cite{Gamberg:2013kla} an even larger error band was obtained for the Sivers-type contribution by using the so-called ``scan procedure''~\cite{Anselmino:2012rq}.  Unlike ours, this band ends up crossing the $x_F$-axis at all values of $x_F > 0$ considered.\footnote{We also mention that the Sivers function used in Ref.~\cite{Gamberg:2013kla} has a flavor-dependent large-$x$ behavior.  The only noticeable difference (in addition to the much larger error band) is a slower decrease in $A_N^\gamma$ at large $x_F$ as compared to our curves.  Thus, allowing for such a feature at large $x$ does not alter our conclusions.}  Therefore, one most likely cannot claim that there exists a robust prediction of a nonzero $A_N^\gamma$ at any region of $x_F > 0$.  In order to pin down the observable more exactly, and in general $A_N$ for any process in the forward region that relies on TMD inputs, one needs more information on TMDs at larger $x$.  This is one of the goals of the 12 GeV upgrade at Jefferson Lab~\cite{Dudek:2012vr}.  Finally, in Fig.~\ref{fixxf} we give $A_N^\gamma$ vs.~$q_T$ at a fixed $x_F = 0.25$ for $\sqrt{S} = 200,\,510\,{\rm GeV}$.  We see the same trend as Fig.~\ref{fixeta} but not as fast a falloff.  Recall that the twist-3 calculation of $A_N^{\pi^0}$ also shows a similar slow decrease as a function of pion transverse momentum~\cite{Kanazawa:2010au,Kanazawa:2014dca}, which is consistent with the observed data~\cite{Adams:2003fx,Adamczyk:2012xd,HEPPELMANN:2013ewa}. The asymmetry is $\sim \!-(1-2\%)$, and in the 200 GeV case the SFP piece can give some contribution at higher-$q_T$.  Also, there seems to be no dependence on $\sqrt{S}$ in this scenario.  Overall, from this numerical study we see the dominance of the chiral-even SGP term could allow one to ``cleanly'' extract the Qiu-Sterman function $G_F(x,x)$ and resolve the sign-mismatch crisis as well as comment on the process dependence of the Sivers function.  It is also important to mention that $A_N^\gamma$ has been studied in the GPM, where one finds a {\it positive} asymmetry~\cite{D'Alesio:2004up, Anselmino:2013rya}.  Thus, a measurement of $A_N^\gamma$ could distinguish between the GPM and twist-3 approaches.
%
%
\section{Summary and conclusion} \label{s:sum}

In this paper we have considered the TSSA in direct photon production from proton-proton collisions.  In particular, we have calculated the twist-3 contribution to
$A_N^\gamma$ that originates from quark-gluon-quark correlations inside
the unpolarized proton. This arises as the coupling of the twist-3 chiral-odd function $E_F(x_1,x_2)$ with the twist-2 chiral-odd transversity.  Both SGPs and SFPs were computed.  Combined with the twist-3 effects
inside the transversely polarized proton~\cite{Qiu:1991wg,Kouvaris:2006zy,Ji:2006vf,Koike:2006qv,Kanazawa:2011er,Gamberg:2012iq,Kanazawa:2012kt,Gamberg:2013kla,Koike:2011nx}, whose results we summarized,
this completes all twist-3 multiparton correlation effects to the asymmetry.  Moreover, using known TMD inputs that have relations to the twist-3 functions that arise in the analytical formulas, we provided a numerical study of this full solution.  We found that the piece involving chiral-odd functions is negligible (with the SFP term actually vanishing identically in the analytical calculation), and also that the SFP part of the term involving chiral-even functions is small.  This leaves the entire effect as due to the ``Sivers-type'' Qiu-Sterman mechanism.  Using the Sivers function extracted from SIDIS~\cite{Anselmino:2008sga}, our results, including uncertainties, show that $A_N^\gamma$ could be on the order of several (negative) percent in the forward region. Given that tri-gluon contributions to $A_N^\gamma$ are extremely small in this $x_F$-regime~\cite{Koike:2011nx}, we believe this observable could provide a ``clean'' extraction of the Qiu-Sterman function $G_F(x,x)$.  In light of the ``sign-mismatch'' crisis involving $G_F(x,x)$ and the Sivers function~\cite{Kang:2011hk}, and the recently proposed solution to this issue that relies on the twist-3 fragmentation mechanism~\cite{Kanazawa:2014dca}, such an extraction is of vital importance. In addition, one can obtain important information on the process dependence of the Sivers function as well as help discriminate between the GPM and twist-3 formalisms.  Thus, measurements of processes that do not have fragmentation contributions are crucial.  Already at RHIC the A$_N$DY Collaboration has measured the jet TSSA~\cite{Bland:2013pkt}, and the PHENIX Collaboration~\cite{PHENIX:BeamUse} and the STAR Collaboration~\cite{STAR:BeamUse} have planned experiments for the direct photon TSSA analyzed here. 

%
%
\section*{Acknowledgments}

We thank the authors of Ref.~\cite{Anselmino:2008sga}, especially A.~Prokudin, for providing us with the parameter sets of the Sivers
function that allowed us to draw the error bands in our figures. 
We also appreciate useful exchanges with M.~Chiu and A.~Vossen about the kinematics that will be accessed by the PHENIX and STAR experiments, respectively, when measuring $A_N^\gamma$.  This work has been supported by the Grant-in-Aid for
Scientific Research from the Japanese Society of Promotion of Science
under Contract No.~26287040 (Y.K.), the National Science
Foundation under Contract No.~PHY-1205942 (K.K. and A.M.), and the RIKEN BNL
Research Center (D.P.).

%
%
%
%
\section*{Appendix A: Derivation of the master formula in Eq.~(\ref{relation})}

In this appendix we will provide a derivation of Eq.~(\ref{relation}) for the SGP contribution.
For the quark-gluon-quark correlation function $G_F(x,x)$,
it has been shown that the hard cross section for the SGP contribution can be related to 
the twist-2 unpolarized cross section.
Here we shall show that the SGP cross section for the
chiral-odd correlation function $E_F(x,x)$ can also be related to a certain
cross section for 2-to-2 parton scattering, 
but in a slightly different form as shown in (\ref{relation}). 
To this end we first define the hard part $H^\alpha (k_1,k_2)$
for $E_F(x_1,x_2)$ by pulling out the factor 
$\gamma_5\pslash \gamma_\kappa\epsilon^{\kappa \alpha np}$
from $M_F^\alpha(x_1,x_2)$ to $S_\lambda (k_1,k_2,x'p',q)p^\lambda $
and taking the Dirac-traces,
\begin{eqnarray}
 H^\alpha(k_1,k_2) = {\rm Tr} \left[ S_\lambda(k_1,k_2,x'p',q) p^\lambda\gamma_5\Slash{p}
			       \gamma_\kappa \epsilon^{\kappa\alpha
			       np}\right],
\end{eqnarray}
where we have suppressed $x'p'$ and $q$ in the argument of $H^\alpha$ for short.  Multiplying $\left.{\partial 
H^\alpha(k_1,k_2) / \partial k_2^\alpha}\right|_{{\rm c.l.}}^{\rm pole}$ by
$x/4$ and taking an appropriate color trace,
one obtains the cross section in Eq.~(\ref{e:co_result}).
Corresponding to the two diagrams in Fig. 2, 
one can write $H^\alpha(k_1,k_2)$ as the sum of two contributions,
\beq
H^\alpha (k_1,k_2) = H_L^\alpha (k_1,k_2)+H_R^\alpha (k_1,k_2)\,,
\eeq
where $H_L$ and $H_R$ denote the contribution from 2(a) and 2(b), respectively, 
and the momenta $k_1$ and $k_2$ are assigned as shown in Fig.~1(a).
Since the extra coherent gluon line coming out of the unpolarized proton
is attached to the incoming quark lines from the polarized proton, $H_L$ and $H_R$ have the 
following structure:
\begin{align}
H_{L}^\alpha (k_1,k_2) &={\rm Tr}\left[\Fbar_{\mu\nu}(k_2,x'p',q){1\over 2}\gamma_5\Sslash L(k_1,k_2) 
T_A F^{\mu\nu}(k_1,k_2-k_1+x'p',q)
\gamma_5\pslash\gamma_{\kappa}\epsilon^{\kappa\alpha np}\right]\nonumber\\
&\quad\times \delta\!\left( (k_2+x'p'-q)^2\right),
\label{app1}\\
H_{R}^\alpha (k_1,k_2) &={\rm Tr}\left[\Fbar_{\mu\nu}(k_2,k_1-k_2+x'p',q)R(k_1,k_2) {1\over 2}\gamma_5\Sslash  T_A
F^{\mu\nu}(k_1,x'p',q)
\gamma_5\pslash\gamma_{\kappa}\epsilon^{\kappa\alpha np}\right]\nonumber\\
&\quad \times \delta\!\left( (k_1+x'p'-q)^2\right),
\label{app2}
\end{align}
where $F^{\mu\nu}(k_1,k',q)$ represents the hard scattering part denoted by the gray circle
in the left of the cut in Fig.~1(a) with the incoming momenta $k_1$ and $k'$ for the (anti)quark line
entering the circle, $\Fbar_{\mu\nu}(k_1,k',q)= \gamma^0 F_{\mu\nu}(k_1,k',q)^\dagger \gamma^0$, and
$L(k_1,k_2)$ and $R(k_1,k_2)$ represent the factors associated with the barred quark propagator 
which produces the SGP in Fig. 2,
\begin{align}
L(k_1,k_2) &= x'\pslash' \pslash {-1\over \kslash_1 -\kslash_2- x'\pslash' +i\epsilon},
\label{appL}\\
R(k_1,k_2) &= {-1\over \kslash_2 -\kslash_1- x'\pslash' -i\epsilon} x'\pslash \pslash'.
\label{appR}
\end{align}
Here we remark that the polarization tensor
$-g_{\mu\nu}$ has been used in (\ref{app1}) and (\ref{app2}) for the
final photon and the gluon.  This procedure is allowed following the same
argument presented in Sec.~3.1.3 of~\cite{Eguchi:2006mc}.
In (\ref{app1}) and (\ref{app2}), we explicitly inserted the color matrix $T_A$
corresponding to the attachment of the coherent gluon line.
But, for simplicity in the notation, we omit the color index $A$
from $H_{L,R}^\alpha$. 
Following \cite{Koike:2006qv,Koike:2007rq,Koike:2011ns}, we calculate the derivative
$\left.{\partial H^\alpha(k_1,k_2) / \partial k_2^\alpha}\right|_{{\rm c.l.}}^{\rm pole}$
by keeping the factors
$F^{\mu\nu}$ and $\Fbar_{\mu\nu}$ intact.  We note that
the derivative $\partial/\partial k_2^\alpha$ hits either $L(k_1,k_2)$ ($R(k_1,k_2)$) or other factors
in $ H^\alpha(k_1,k_2)$.  
One thus needs
\begin{align}
L(x_1p,x_2p)&={1\over x_2-x_1+i\epsilon}x'\pslash',
\label{Lcl}\\
R(x_1p,x_2p)&={-1\over x_2-x_1+i\epsilon}x'\pslash',
\label{Rcl}
\end{align}
and
\begin{align}
\left.{\partial L(k_1,k_2)
\over \partial k_2^\alpha}\right|_{{\rm c.l.}}&={1\over 2 p\cdot p'}{1\over x_2-x_1+i\epsilon}\,\pslash'\pslash\gamma_\alpha
+ {(- p'_\alpha) \over p\cdot p'}\left( {1 \over x_2-x_1+i\epsilon}\right)^2 x' \pslash',
\label{derivL}\\
\left.{\partial R(k_1,k_2)
\over \partial k_2^\alpha}\right|_{{\rm c.l.}}&={1\over 2 p\cdot p'}{1\over x_2-x_1+i\epsilon}\, \gamma_\alpha \pslash\pslash'
+ {(p'_\alpha) \over p\cdot p'}\left( {1 \over x_2-x_1+i\epsilon}\right)^2 x' \pslash'.
\label{derivR}
\end{align}
For convenience
we decompose 
$\left.{\partial H^\alpha(k_1,k_2) / \partial k_2^\alpha}\right|_{{\rm c.l.}}^{\rm pole}$ into three pieces:~(i) the double pole terms coming from the
second terms in (\ref{derivL}) and (\ref{derivR}) when the derivative hits $L$ or $R$; 
(ii) the simple pole terms coming from the
first terms in (\ref{derivL}) and (\ref{derivR}); and
(iii) the simple pole terms which occur with (\ref{Lcl}) and (\ref{Rcl})
when the derivative hits other factors in $H^\alpha(k_1,k_2)$.  

The pole part of the contribution (i) reads
\beq
&&\left.{\partial H^\alpha(k_1,k_2) \over \partial k_2^\alpha}\right|^{({\rm i})}_{{\rm c.l.}}\nonumber\\
&&={-p'_\alpha\over p\cdot p'} \left[\left( {1\over x_2-x_1 +i\epsilon}\right)^2\right]^{\rm pole}
{\rm Tr} \Big[ \Fbar_{\mu\nu}(x_2 p,x'p',q){1\over 2}\gamma_5 \Sslash 
x'\pslash' T_A F^{\mu\nu}(x_1p,(x_2-x_1)p+x'p',q)\nonumber\\
&&\qquad\qquad\qquad\qquad\qquad\qquad\qquad \times\,
\gamma_5\pslash\gamma_\kappa\epsilon^{\kappa\alpha np}\Big]
\delta\!\left( (x_2p+x'p'-q)^2\right)\nonumber\\
&&+\,{p'_\alpha\over p\cdot p'} \left[\left( {1\over x_2-x_1 +i\epsilon}\right)^2\right]^{\rm pole}
{\rm Tr} \Big[ \Fbar_{\mu\nu}(x_2 p,(x_1-x_2)p +
x'p',q){1\over 2}\gamma_5 \Sslash x'\pslash' T_A F^{\mu\nu}(x_1p,x'p',q)\nonumber\\
&&\qquad\qquad\qquad\qquad\qquad\qquad\qquad \times \,
\gamma_5\pslash\gamma_\kappa\epsilon^{\kappa\alpha np}\Big]
\delta\!\left( (x_1p+x'p'-q)^2\right).
\eeq
Owing to the presence of the factor $\delta'(x_2-x_1)$, we need to keep terms up to linear order in
$x_2-x_1$ in the Taylor expansion of the right-hand-side with respect to $x_2$ around $x_1$.
In this expansion, the double pole terms cancel and the remaining simple pole terms
can be combined into a compact form by transforming the derivative with respect to $x_2p$
into that with respect to $x'p'$.  
One thus obtains
\beq
&&\left.{\partial H^\alpha(k_1,k_2) \over \partial
k_2^\alpha}\right|^{({\rm i})}_{{\rm c.l.}} \nonumber\\
&&
= {-p'_\alpha \over p\cdot p'}\left( {1\over x_2-x_1+i\epsilon}\right)^{\rm pole}\nonumber\\
&&\times \left\{
p^\lambda{\partial \over \partial x'p'^\lambda}
{\rm Tr}\left[
\Fbar_{\mu\nu}(x_1p,x'p',q){1\over 2}\gamma_5 \Sslash x'\pslash' T_A F^{\mu\nu}(x_1p,x'p',q)
\gamma_5\pslash\gamma_\kappa\epsilon^{\kappa\alpha np}
\delta\!\left( (x_1p+x'p'-q)^2\right)\right]\right.\nonumber\\
&&\left.
\hspace{1cm}-\, {\rm Tr}\left[
\Fbar_{\mu\nu}(x_1p,x'p',q){1\over 2}\gamma_5 \Sslash \pslash T_A F^{\mu\nu}(x_1p,x'p',q)
\gamma_5\pslash\gamma_\kappa\epsilon^{\kappa\alpha np}
\delta\!\left( (x_1p+x'p'-q)^2\right)\right]
\right\}.
\label{cont1}
\eeq
Similarly, one obtains for the contributions (ii) and (iii)
\beq
&&\left.{\partial H^\alpha(k_1,k_2) \over \partial
k_2^\alpha}\right|^{({\rm ii})}_{{\rm c.l.}} \nonumber\\
&&
=\left( {1\over x_2-x_1+i\epsilon}\right)^{\rm pole}{\rm Tr}
\bigg[
\Fbar_{\mu\nu}(x_1p,x'p',q){1\over 2}\gamma_5
\left(
{p\cdot S\over p\cdot p'}
 \gamma_\alpha \pslash' - {p'_\alpha \over p\cdot p'}\Sslash \pslash
-{S_\alpha\over p\cdot p'}\pslash\pslash' +  \Sslash\gamma_\alpha\right) 
\nonumber\\
&&\qquad\qquad\hspace{3cm} \times \,T_A F^{\mu\nu}(x_1p,x'p',q)
\gamma_5\pslash\gamma_\kappa\epsilon^{\kappa\alpha np}
\delta\!\left( (x_1p+x'p'-q)^2\right)\bigg],
\label{cont2}
\eeq
and 
\beq
&&\left.{\partial H^\alpha(k_1,k_2) \over \partial
k_2^\alpha}\right|^{({\rm iii})}_{{\rm c.l.}} \nonumber\\
&&
=\left( {1\over x_2-x_1+i\epsilon}\right)^{\rm pole}
\bigg\{
{\partial \over \partial x'p'^\alpha}
{\rm Tr}\bigg[
\Fbar_{\mu\nu}(x_1p,x'p',q){1\over 2}\gamma_5 \Sslash x'\pslash' T_A F^{\mu\nu}(x_1p,x'p',q)\nonumber\\
&&
\qquad\qquad\qquad\qquad\qquad\qquad\qquad\times\,
\gamma_5\pslash\gamma_\kappa\epsilon^{\kappa\alpha np}
\delta\!\left( (x_1p+x'p'-q)^2\right)\bigg]\nonumber\\
&&
\hspace{0.5cm}-\,{\rm Tr}\bigg[
\Fbar_{\mu\nu}(x_1p,x'p',q){1\over 2}\gamma_5 \Sslash \gamma_\alpha T_A F^{\mu\nu}(x_1p,x'p',q)
\gamma_5\pslash\gamma_\kappa\epsilon^{\kappa\alpha np}
\delta\!\left( (x_1p+x'p'-q)^2\right)\bigg]\bigg\}.
\label{cont3}
\eeq
By taking the sum of (\ref{cont1}), (\ref{cont2}) and (\ref{cont3}), we finally find
\beq
&&\left.{\partial H^\alpha(k_1,k_2) \over \partial
k_2^\alpha}\right|^{{\rm (i)+(ii)+(iii)}}_{{\rm c.l.}} \nonumber\\
&&
= \left( {1\over x_2-x_1+i\epsilon}\right)^{\rm pole}
\nonumber\\
&&\times \bigg\{
\!\left( {\partial \over \partial x'p'^\alpha}
- {p'_\alpha p^\lambda \over p\cdot p'}
{\partial \over \partial x'p'^\lambda}\right)
{\rm Tr}\bigg[
\Fbar_{\mu\nu}(x_1p,x'p',q){1\over 2}\gamma_5 \Sslash x'\pslash' T_A F^{\mu\nu}(x_1p,x'p',q)
\nonumber\\
&&\qquad\qquad\qquad\qquad\qquad \hspace{2cm}\times\,\gamma_5\pslash\gamma_\kappa\epsilon^{\kappa\alpha np}
\delta\!\left( (x_1p+x'p'-q)^2\right)\bigg]\nonumber\\
&&+\, {\rm Tr}\bigg[
\Fbar_{\mu\nu}(x_1p,x'p',q){1\over 2}\gamma_5 
\left(
{p\cdot S\over p\cdot p'}
 \gamma_\alpha \pslash' 
-{S_\alpha\over p\cdot p'}\pslash\pslash' \right)T_A F^{\mu\nu}(x_1p,x'p',q) \nonumber\\
&&\qquad\qquad\qquad
\times\, 
\gamma_5\pslash\gamma_\kappa\epsilon^{\kappa\alpha np}
\delta\!\left( (x_1p+x'p'-q)^2\right)\bigg]
\bigg\}.
\label{master}
\eeq
This formula provides a convenient tool to calculate the SGP cross section
in a general frame.  
In a frame where $p$ and $p'$ are collinear, one has $p\cdot S=0$ and
(\ref{master}) results in Eq.~(\ref{relation}) with 
\begin{equation}
 S (xp,x'p',q) = \Fbar_{\mu\nu}(xp,x'p',q){1\over 2}\gamma_5 \Sslash
  x'\pslash' T_A F^{\mu\nu}(xp,x'p',q) \delta\!\left( (xp+x'p'-q)^2\right).
\end{equation}
Note that one should first keep $p'_\perp \neq 0$
when taking the derivative $\partial /\partial x'p'{^\alpha} $ and then take the limit 
$p'_\perp \to 0$.  The derivative $\partial /\partial x'p'{^\alpha} $
can be taken by using a similar procedure as described in \cite{Koike:2011mb}.


\end{document}